\documentclass[10pt,conference]{IEEEtran}

\usepackage[backend=biber,style=ieee,sorting=none]{biblatex}
\usepackage{graphicx}
\usepackage{amsmath}
\usepackage{booktabs}
\usepackage{array}
\usepackage{xcolor}
\usepackage{tikz}
\usetikzlibrary{arrows.meta, positioning, shapes.geometric, calc, fit, backgrounds}
\usepackage{hyperref}
\usepackage{caption}
\usepackage{microtype}
\usepackage{adjustbox}

\definecolor{intelblue}{RGB}{0, 113, 197}

\title{Single 32-bit Sub-Channel DDR5 DIMMs:\\
  Architecture, Performance Bounds,\\
  and Standardisation}

\author{\IEEEauthorblockN{Chih-Hua Ke}
\IEEEauthorblockA{Product Performance Research Division\\
GIGABYTE Technology Co., Ltd.\\
New Taipei, Taiwan\\
\textit{hicookie@gigabyte.com}}}

\date{May 2026}

\begin{document}

\maketitle

% ── Abstract ───────────────────────────────────────────────────────────────
\begin{abstract}
DDR5 SDRAM partitions each 64-bit memory channel into two independent 32-bit sub-channels. A DIMM populating only one sub-channel halves the die count required for a given module, enabling 8\,GB modules with current 16\,Gbit dies that the standard topology cannot achieve. The configuration has been used by the enthusiast overclocking community since 2021 to set DDR5 frequency world records on three successive Intel platform generations, and has recently received attention as a candidate for cost-reduced volume modules under the contemporaneous DRAM supply constraints. We derive the transaction-width identity grounding the JEDEC sub-channel design: 32-bit $\times$ BL16 transfers exactly one 64-byte x86 cache line per burst. Using a roofline model we quantify performance impact across workload classes (40--60\% throughput degradation in bandwidth-bound workloads, $<$10\% in latency-dominated workloads), and identify a bandwidth inversion at DDR5-4800 below DDR4-3200. Platform analysis shows architectural incompatibility with AMD AM5 as a consequence of the unified 64-bit UMC training model. We further show that the JEDEC SPD specification (JESD400-5D.01) already encodes single sub-channel modules natively in Byte 235, and identify the surrounding ecosystem standardisation gap.
\end{abstract}


% ── I. Introduction ────────────────────────────────────────────────────────
\section{Introduction}

DDR5 SDRAM \cite{jedec_ddr5_2020} partitions each 64-bit memory channel into two independent 32-bit sub-channels (SC0 and SC1), each with its own command/address bus, bank-group hierarchy, and PHY instance. This architecture provides higher access parallelism and per-sub-channel refresh granularity at the cost of a higher minimum die count:  a standard single-rank consumer UDIMM with 16 Gbit $\times$8 dies requires eight data dies, yielding a 16 GB capacity floor; DDR5's mandatory on-die ECC is internal to each die and does not add to the die count. A DIMM populating only Sub-Channel A (SC0) and leaving Sub-Channel B (SC1) unpopulated halves the die count and the data-bus width, enabling 8\,GB modules at reduced bill-of-materials cost.

Industry interest in this configuration as a candidate for volume-tier modules has increased over 2025--2026 in the context of the DRAM supply constraints associated with AI-related wafer reallocation \cite{trendforce_ai_dram_2025, idc_memory_2026}: HBM consumes approximately three times the wafer area per bit of equivalent commodity DDR5 \cite{tomshw_hbm_2025}, AI applications are projected to consume approximately 20\% of global DRAM wafer output in 2026 \cite{trendforce_ai_dram_2025}, and the corresponding contraction in commodity DDR4 production has reduced the availability of an alternative cost-baseline for 8\,GB modules \cite{trendforce_ddr4_2025, taipeitimes_nanya_2026}. The JEDEC DDR5 specification \cite{jedec_ddr5_2020} pairs a 32-bit per-sub-channel data path with a doubled burst length (BL16, vs BL8 in DDR4) so that one burst on a single sub-channel transfers exactly one 64-byte x86 cache line, preserving per-transaction access granularity. A consequence is that a single populated sub-channel is sufficient to satisfy a CPU cache fill in one burst transaction; this property underpins the operational behaviour of single sub-channel modules.

The 32-bit single sub-channel configuration is not architecturally new: it has been used by the enthusiast overclocking community since the launch of Intel Alder Lake (12th Gen Core, late 2021) to set the publicly recognised DDR5 memory frequency world record across three successive Intel platform generations \cite{gigabyte_ddr5_10022_2022, gigabyte_ddr5_11618_2023, gigabyte_ddr5_13530_2025}. The configuration is therefore well-characterised at the electrical level across a wide range of operating conditions, including extreme overclocking, on contemporary Intel platforms. The 2025--2026 productisation interest is a repositioning of an existing technique toward cost-reduced consumer modules, not a new architectural proposal.

This paper provides a rigorous mathematical derivation of the cache-line-granularity sufficiency property, analyses its scope and limitations, quantifies the performance trade-offs across workload classes using the roofline model, evaluates platform compatibility with the Intel iMC and AMD UMC, and characterises the JEDEC standardisation status of the configuration.

The contributions of this work are:
\begin{enumerate}
  \item A formal derivation of DDR5 sub-channel transaction width and its equivalence to the x86 cache line size.
  \item A roofline-based workload classification of performance impact, with explicit identification of a bandwidth inversion at DDR5-4800 below DDR4-3200.
  \item A comparative architectural analysis of Intel iMC and AMD UMC sub-channel handling, identifying the root cause of platform incompatibility.
  \item Identification that the JEDEC SPD specification already encodes the single sub-channel module class (Byte 235 bits 7--5 = \texttt{000}) natively, and characterisation of the surrounding ecosystem standardisation gap.
  \item A discussion of the empirical foundation for the configuration provided by published HWBOT-verified frequency records on three Intel platform generations.
\end{enumerate}

% ── II. Related Work ───────────────────────────────────────────────────────
\section{Related Work}

Prior literature on low-cost DRAM module design has largely focused on LPDDR and mobile DRAM cost optimisation. Malladi \textit{et al.} \cite{malladi2012dram} demonstrate the energy and density advantages of mobile DRAM for datacenter applications, establishing a precedent for non-standard module configurations in cost-sensitive deployments. Micron's DDR5 overview \cite{micron_ddr5_2022} discusses the sub-channel architecture from a device perspective but does not evaluate asymmetric population. Patterson and Hennessy \cite{patterson_hennessy_2020} provide the roofline model framework applied in Section~\ref{sec:perf}.

Vendor-specific commercial implementations of single sub-channel DDR5 modules have appeared in 2026 under various trade names; this paper analyses the underlying architecture independent of any specific commercial implementation, and uses the JEDEC sub-channel terminology. The contemporaneous DRAM supply context has been documented in industry analysis \cite{trendforce_ai_dram_2025, idc_memory_2026, taipeitimes_nanya_2026, trendforce_ddr4_2025}. To the authors' knowledge, no prior peer-reviewed work has formally characterised single sub-channel DDR5 module viability, performance bounds, or platform constraints; this paper addresses that gap.

The absence of prior analysis reflects three structural factors. First, JEDEC standardisation processes have historically defined module classes by capacity and speed rather than by sub-channel population, and the standard provides no module-identification machinery for half-width modules even though the SPD encoding for them exists (Section~\ref{sec:jedec_spd_field}). Second, module vendors avoid non-standard SKUs whose validation and platform-compatibility burdens are difficult to amortise without a defined reference platform. Third, the asymmetric platform support landscape between Intel and AMD has discouraged industry-wide attention. Intel's iMC architecture documented in the Alder Lake datasheet \cite{intel_ddr5_2021} together with the 2025--2026 supply pressures provided the conditions under which this configuration has received commercial attention.

% ── III. DDR5 Sub-Channel Architecture ────────────────────────────────────
\section{DDR5 Sub-Channel Architecture}

\subsection{Channel Partitioning at the JEDEC Level}

The JEDEC DDR5 specification \cite{jedec_ddr5_2020} defines a 64-bit data bus per channel, partitioned into two independent 32-bit sub-channels. Each sub-channel contains:
\begin{itemize}
  \item A dedicated 32-bit data path
  \item An independent command/address (CA) bus
  \item Independent row/column decoders
  \item 16 banks organised into 8 bank groups of 2 banks each
\end{itemize}

The two sub-channels of a JEDEC DDR5 channel are electrically and logically independent: each may issue simultaneous READ/WRITE commands to different rows without inter-sub-channel coordination. This is a departure from DDR4, where the 64-bit bus was monolithic and a single command bus controlled all 64 data lines. The Micron DDR5 white paper \cite{micron_ddr5_2022} describes the design intent: doubling the burst length from BL8 (DDR4) to BL16 enables a sub-channel partition that preserves the 64-byte-per-transaction granularity required by x86 cache lines while doubling channel concurrency.

\subsection{Physical Realisation: DIMM Slot Topology}
\label{sec:slot_topology}

The mapping between JEDEC sub-channels and physical DIMM slots is platform-specific. Two distinct topologies are relevant for the present analysis.

In Intel platforms through 13th Gen Core (Raptor Lake) and 14th Gen Core (Raptor Lake Refresh), the two sub-channels of a given memory controller channel were both routed to the same physical DIMM slot. A standard DIMM in that slot exposed both SC0 and SC1 to its memory controller, and a cost-reduced ``single sub-channel'' DIMM populated only the DRAM dies corresponding to one of the two sub-channels (typically SC0), leaving the other half of the module unpopulated. The Intel iMC was observed to operate correctly in this asymmetric configuration in overclocking practice (Section~\ref{sec:oc_priority}).

Intel Arrow Lake (Core Ultra 200S, 2024) introduces a different physical topology. Each memory controller (CMC0, CMC1) drives two physically separate DIMM slots, and the two 32-bit sub-channels of a given channel are routed to different slots: in a 2-DIMM-per-CMC layout, slot~A receives the SC0 trace and slot~B receives the SC1 trace from the same CMC. Figure~\ref{fig:imc_umc} (Section~\ref{sec:platform}) illustrates this. Under this topology, populating only one slot exposes only one sub-channel of each CMC to the host, regardless of how the DIMM itself is internally constructed. The cost-reduced single sub-channel module class, in this context, is realised as a DIMM whose physical width corresponds to a single 32-bit sub-channel.

The architectural and performance analysis in Sections~\ref{sec:fullfill}--\ref{sec:perf} is conducted at the JEDEC sub-channel layer and is therefore invariant to which of these two physical topologies is in use. Platform compatibility analysis (Section~\ref{sec:platform}) is platform-specific; we focus on the Arrow Lake topology because it is the most recent Intel client platform and corresponds to the reference platform used for the most recent of the world records discussed in Section~\ref{sec:oc_priority}.

\subsection{Transaction Width and Burst Length}

DDR5 mandates Burst Length 16 (BL16) as the standard operating mode \cite{jedec_ddr5_2020, micron_ddr5_2022}. A BL16 transaction transfers data across 16 consecutive half-clock edges (8 rising, 8 falling); the DDR factor is implicit in the burst count. The transaction width per sub-channel is:

\begin{equation}
T_{SC} = \frac{W_{\text{bus}}}{8} \times BL = \frac{32}{8} \times 16 = 64\,\text{bytes}.
\label{eq:txsc}
\end{equation}

For comparison, DDR4 \cite{jedec_ddr4_2020} with a 64-bit bus and BL8 delivers:

\begin{equation}
T_{\text{DDR4}} = \frac{64}{8} \times 8 = 64\,\text{bytes}.
\label{eq:txddr4}
\end{equation}

Both generations yield identical transaction widths per physical command. A fully populated DDR5 channel firing both sub-channels simultaneously achieves 128\,bytes per burst --- doubling the per-command transfer relative to DDR4 --- but each individual sub-channel remains a 64-byte unit.

\subsection{Full Configuration Comparison}

Table~\ref{tab:full_picture} enumerates transaction widths and theoretical peak bandwidths across all relevant configurations. A notable observation is that two single sub-channel DIMMs in a dual-channel topology produce peak bandwidth identical to one standard DDR5 DIMM at the same data rate (44.8\,GB/s at DDR5-5600). This offers a cost-disaggregated path to standard effective bandwidth: two cheap modules can substitute for one standard module while preserving aggregate throughput, at the expense of consuming both DIMM slots.

\begin{table*}[t]
  \centering
  \caption{Transaction Width and Peak Bandwidth Across DDR4 and DDR5 Configurations}
  \label{tab:full_picture}
  \small
  \begin{tabular}{llccccrr}
    \toprule
    \textbf{Standard} & \textbf{Configuration} & \textbf{DIMMs} & \textbf{Bus} & \textbf{BL} & \textbf{Tx Width} & \textbf{Speed} & \textbf{BW (GB/s)} \\
    \midrule
    DDR4 \cite{jedec_ddr4_2020} & 1 DIMM, 1 channel            & 1 & 64-bit  &  8 & \textbf{64\,B}  & DDR4-3200 & 25.6 \\
    DDR4                        & 2 DIMM, 2 channel (dual)     & 2 & 128-bit &  8 & \textbf{128\,B} & DDR4-3200 & 51.2 \\
    \midrule
    DDR5 \cite{jedec_ddr5_2020} & 1 DIMM, 1ch, dual SC (std.) & 1 & 64-bit  & 16 & \textbf{128\,B} & DDR5-5600 & 44.8 \\
    DDR5                        & 2 DIMM, 2ch, dual SC (std.) & 2 & 128-bit & 16 & \textbf{256\,B} & DDR5-5600 & 89.6 \\
    \midrule
    DDR5 (this work)            & 1 DIMM, 1ch, single SC      & 1 & 32-bit  & 16 & \textbf{64\,B}  & DDR5-5600 & 22.4 \\
    DDR5 (this work)            & 2 DIMM, 2ch, single SC      & 2 & 64-bit  & 16 & \textbf{128\,B} & DDR5-5600 & 44.8 \\
    \bottomrule
    \multicolumn{8}{l}{\footnotesize Tx Width = (bus\_bits/8) $\times$ BL. BW = (bus\_bits/8) $\times$ MT/s.} \\
    \multicolumn{8}{l}{\footnotesize x86 cache line = 64\,bytes. All configurations transfer one cache line per burst transaction.} \\
  \end{tabular}
\end{table*}

% ── IV. Cache-Line-Granularity Sufficiency ────────────────────────────────
\section{Cache-Line-Granularity Sufficiency of a Single Sub-Channel}
\label{sec:fullfill}

\subsection{x86 Cache Line Granularity}

Modern x86 processors universally employ a 64-byte cache line as the fundamental unit of transfer between the cache hierarchy and DRAM \cite{patterson_hennessy_2020}. A last-level cache (LLC) miss generates a demand fill of exactly 64\,bytes; the memory subsystem must satisfy this request atomically --- a cache line fill. This 64-byte granularity has been architecturally invariant across Intel and AMD platforms for over two decades, from Pentium 4 through current Alder/Meteor/Arrow Lake and Zen 4/5.

\subsection{Sub-Channel Sufficiency}

Substituting the standard cache line size into Equation~(\ref{eq:txsc}):

\begin{equation}
T_{SC} = 64\,\text{bytes} = L_{x86},
\label{eq:fullfill}
\end{equation}

where $L_{x86} = 64\,\text{bytes}$ denotes the x86 cache line size. The identity (\ref{eq:fullfill}) is in fact the design intent of the DDR5 sub-channel structure as documented in the JEDEC specification \cite{jedec_ddr5_2020}: with BL16 doubled from DDR4's BL8, a 32-bit sub-channel transfers exactly one 64-byte x86 cache line per burst, preserving the per-transaction granularity of DDR4 while enabling per-sub-channel scheduling. A consequence of this property is that a cache miss on a host using a single 32-bit sub-channel can be resolved in a single DRAM burst transaction, without partial fills, split transactions, or pipeline stalls. This holds \textbf{with respect to per-transaction atomicity}; it is narrowly scoped, and does not by itself bound the system-level performance impact of halving the sub-channel count, which is the subject of Section~\ref{sec:perf}.

\subsection{Latency Implications}

Cache miss latency is not degraded by reducing the sub-channel count from two to one. Memory access latency ($t_{RCD} + t_{CL} + t_{RP}$) is a function of DRAM timing parameters, not bus width. A single sub-channel DIMM at DDR5-5600 exhibits the same first-access latency as a dual sub-channel DIMM at the same speed. We note that DDR5 does not inherently improve absolute latency over DDR4 --- both generations operate at approximately 14\,ns first-access latency at their respective standard speeds. The latency profile of single sub-channel DDR5 versus standard DDR5 is therefore neutral, not positive.

\subsection{Empirical Use in the Overclocking Community}
\label{sec:oc_priority}

The 32-bit single sub-channel configuration has been used by the enthusiast overclocking community since the launch of Intel Alder Lake (12th Gen Core, late 2021) as a frequency-headroom technique: with one sub-channel disabled and the corresponding DQ/DQS/CA loading removed from the bus, the remaining sub-channel exhibits reduced capacitive loading at the iMC's PHY interface and supports per-pin data rates beyond those reported on fully populated 64-bit channels at the same generation. The configuration has been used to set the publicly recognised DDR5 memory frequency world record across three successive Intel platform generations. The records are independently verified by HWBOT, which is the principal third-party records authority for competitive overclocking and which publishes full hardware configuration disclosure for each submission:

\begin{itemize}
  \item \textbf{Alder Lake (May 2022).} DDR5-10022 (5011\,MHz) on Z690 AORUS TACHYON with Core i9-12900K, set by HiCookie (the present author) under LN$_2$ cooling \cite{gigabyte_ddr5_10022_2022, hwbot_5379_2022}.
  \item \textbf{Raptor Lake Refresh (October 2023).} DDR5-11618 (5809.2\,MHz) on Z790 AORUS TACHYON X with Core i9-14900K, set by HiCookie at IEM 2023 Sydney \cite{gigabyte_ddr5_11618_2023, hwbot_5376447_2023}.
  \item \textbf{Arrow Lake (November 2025).} DDR5-13530 on Z890 AORUS TACHYON ICE with Core Ultra 200S, set jointly by Sergmann and HiCookie \cite{gigabyte_ddr5_13530_2025, hwbot_5929126_2025}.
\end{itemize}

In each case standard dual sub-channel modules were used, with the second sub-channel disabled at the BIOS / memory reference code level rather than via SPD encoding; this exercises the same iMC code path that an SPD-declared single sub-channel module would invoke at initialisation, but is configured at the platform layer. Each record corresponds to a per-pin data rate that the prevailing 64-bit channel topology of its generation could not sustain. The Arrow Lake record was the subject of an Intel-produced documentary featuring Intel's Chief Overclocking Architect \cite{intel_ddr5_13530_video_2025}. These records establish that the iMC's single sub-channel operating mode is well-characterised at the electrical level across the three most recent Intel client platforms; they do not establish a performance argument for the volume-tier productisation discussed in the remainder of this paper, since the records are set under non-production cooling and timing conditions, and they do not directly validate the SPD-encoded module path that productisation would use.

\subsection{Scope Limitations}

Three important qualifications bound the per-transaction sufficiency claim of Section~\ref{sec:fullfill}:

\textbf{(1) Request concurrency.} A standard dual sub-channel DIMM can service two independent cache line requests simultaneously --- one per sub-channel. A single sub-channel module serialises all cache fill traffic through one 32-bit command queue, halving peak concurrent fill throughput.

\textbf{(2) Prefetch efficiency.} DDR5 optionally supports BL32, delivering:
\begin{equation}
T_{BL32} = \frac{32}{8} \times 32 = 128\,\text{bytes} = 2 \times L_{x86}.
\end{equation}
Dual sub-channel DDR5 at BL16 achieves this natively; single sub-channel requires BL32 to match, which is not the default mode and may incur refresh-interval penalties.

\textbf{(3) Queueing latency under load.} While first-access latency is bus-width-independent, average access latency grows with request queue depth. We estimate 15--25\% higher average memory access latency under typical desktop workloads at bus utilisation exceeding 60\%, due to increased Head-of-Line blocking on the single command queue.

% ── V. Performance Analysis ───────────────────────────────────────────────
\section{Performance Analysis}
\label{sec:perf}

\subsection{Roofline Model Framework}

Following the roofline model \cite{patterson_hennessy_2020}, a workload with arithmetic intensity $I$ (FLOP/byte) operating on a system with peak compute $P$ (FLOP/s) and peak memory bandwidth $B$ (byte/s) is bandwidth-limited when:

\begin{equation}
I < \frac{P}{B}.
\label{eq:roofline}
\end{equation}

Halving $B$ from $B_0$ to $B_0/2$ proportionally compresses the bandwidth-limited region, raising the crossover threshold at which workloads transition from memory-bound to compute-bound operation. Workloads already operating in the compute-bound regime experience no performance impact; those in the memory-bound regime experience throughput degradation proportional to the bandwidth reduction.

\subsection{Effective Throughput}

DDR5's improved bank-group organisation and per-sub-channel scheduling yield higher bus efficiency than DDR4 (estimated 80--90\% vs. 70--80\%). The effective single sub-channel throughput is therefore:

\begin{equation}
B_{\text{eff}} \approx 22.4 \times 0.85 \approx 19\,\text{GB/s},
\end{equation}

comparable to DDR4-3200 single-channel effective throughput ($\approx$18--20\,GB/s, assuming DDR4 bus efficiency of 70--80\%). For latency-dominated workloads operating well below bandwidth saturation, this equivalence is practically sufficient. The 0.85 efficiency factor is an estimate; published DDR5 controller analyses report sustained efficiencies in the 80--90\% range under random-access traffic, with DDR4 typically 70--80\% at comparable utilisation \cite{micron_ddr5_2022}.

\subsection{Peak Bandwidth Across Speeds}

Table~\ref{tab:bandwidth} compares dual and single sub-channel DDR5 at representative speeds. A bandwidth inversion appears at the DDR5-4800 tier: single sub-channel DDR5-4800 delivers 19.2\,GB/s peak (approximately 16\,GB/s effective at 85\% efficiency), below DDR4-3200's approximately 21\,GB/s effective. A migration from DDR4-3200 to single sub-channel DDR5-4800 therefore offers no bandwidth gain; the DDR5 upgrade provides a bandwidth advantage only at the 5600\,MT/s tier and above. For OEM deployments where the DDR5 data rate is locked to the JEDEC base speed by platform configuration, this means single sub-channel modules can yield lower sustained throughput than the DDR4 systems they replace.

A second observation: single sub-channel DDR5-6400 delivers the same peak bandwidth as DDR4-3200 (25.6\,GB/s), so even at the high end of the speed range the bandwidth uplift over DDR4 is marginal. DDR5's value at this tier is in lower operating voltage (1.1\,V vs. 1.2\,V) and improved scheduling efficiency, not raw throughput.

\begin{table*}[!t]
  \centering
  \caption{Peak and Effective Bandwidth: Dual-SC vs. Single-SC DDR5 at Representative Speeds}
  \label{tab:bandwidth}
  \small
  \setlength{\tabcolsep}{6pt}
  \begin{tabular}{@{}lccrr@{}}
    \toprule
    \textbf{Configuration} & \textbf{Speed} & \textbf{Bus} & \textbf{Peak (GB/s)} & \textbf{Effective$^\ddagger$ (GB/s)} \\
    \midrule
    Dual SC (std.)        & DDR5-4800 & 64-bit & 38.4 & $\approx$33 \\
    Dual SC (std.)        & DDR5-5600 & 64-bit & 44.8 & $\approx$38 \\
    Dual SC (std.)        & DDR5-6400 & 64-bit & 51.2 & $\approx$44 \\
    \midrule
    Single SC (this work) & DDR5-4800 & 32-bit & 19.2 & $\approx$16 \\
    Single SC (this work) & DDR5-5600 & 32-bit & 22.4 & $\approx$19 \\
    Single SC (this work) & DDR5-6400 & 32-bit & 25.6 & $\approx$22 \\
    \midrule
    DDR4-3200 (reference) & DDR4-3200 & 64-bit & 25.6 & $\approx$21 \\
    \bottomrule
    \multicolumn{5}{@{}l}{\footnotesize $^\ddagger$Effective bandwidth $\approx$ peak $\times$ 0.85 (DDR5) or peak $\times$ 0.80 (DDR4).} \\
  \end{tabular}
\end{table*}

\subsection{Workload Classification}

Table~\ref{tab:workloads} classifies representative workloads by memory bandwidth sensitivity and estimated performance deficit relative to standard dual sub-channel DDR5-5600. Deficit ranges are derived from roofline model projections and calibrated against published DDR4 vs. DDR5 benchmark deltas \cite{micron_ddr5_2022}.

\begin{table}[!htbp]
  \centering
  \caption{Performance Impact by Workload vs. Standard DDR5-5600}
  \label{tab:workloads}
  \small
  \setlength{\tabcolsep}{4pt}
  \begin{tabular}{@{}lll@{}}
    \toprule
    \textbf{Workload} & \textbf{BW Sens.} & \textbf{Deficit} \\
    \midrule
    Web / office / productivity     & Low       & 2--8\%  \\
    Gaming, GPU-bound               & Low       & 5--12\% \\
    Gaming, CPU-bound$^\dagger$     & Medium    & 15--35\% \\
    Software development            & Low--Med  & 8--15\% \\
    Video playback                  & Low       & 2--5\%  \\
    Video transcoding (x264/x265)   & High      & 30--50\% \\
    iGPU, 1080p+                    & High      & 35--55\% \\
    CPU AI / LLM serving            & High      & 40--60\% \\
    Scientific simulation (HPC)     & High      & 40--60\% \\
    POS / kiosk / embedded          & Very Low  & $<$3\%  \\
    \bottomrule
    \multicolumn{3}{@{}l}{\footnotesize $^\dagger$High-fps titles ($>$240\,fps); CPU-side pressure dominant.} \\
  \end{tabular}
\end{table}

Integrated graphics warrants particular attention. In laptop platforms, where cost-reduced DDR5 is most likely to be deployed, the iGPU shares the system memory bandwidth budget with the CPU. At 1080p, typical iGPU bandwidth requirements range from 18 to 22\,GB/s. Single sub-channel DDR5-5600 has a peak of 22.4\,GB/s and a sustained value of approximately 19\,GB/s at 85\% bus efficiency, both of which fall within or below this iGPU demand range, leaving little margin for concurrent CPU traffic. At single sub-channel DDR5-6400, the sustained value of approximately 22\,GB/s sits at the upper edge of the iGPU demand range and remains insufficient for sustained mixed workloads. In typical mixed iGPU + CPU operation, the controller must time-multiplex demand against a budget that is fully occupied by GPU traffic alone, producing measurable frame-time variance and queueing latency. Single sub-channel modules are therefore not appropriate for laptop deployments with 1080p or higher sustained graphics targets within the consumer DDR5 speed envelope considered.

% ── VI. Platform Compatibility ────────────────────────────────────────────
\section{Platform Compatibility}
\label{sec:platform}

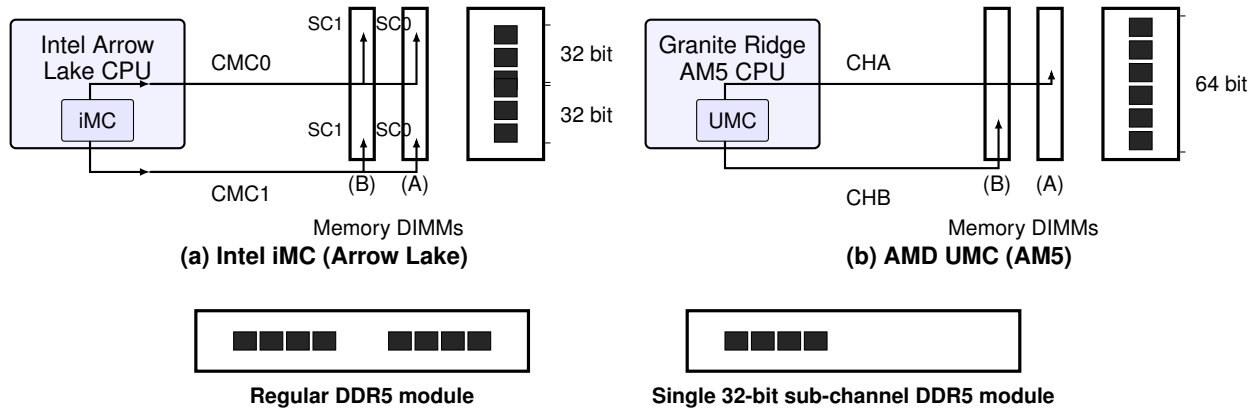
\begin{figure*}[!t]
  \centering
  \begin{tikzpicture}[
      font=\sffamily\small,
      cpu/.style={draw, rounded corners=2pt, thick, minimum width=2.0cm, minimum height=1.7cm, align=center, fill=blue!5},
      mc/.style={draw, rounded corners=1pt, minimum width=0.95cm, minimum height=0.55cm, align=center, fill=blue!10, font=\sffamily\footnotesize},
      slot/.style={draw, very thick, minimum width=0.32cm, minimum height=2.0cm, fill=white},
      dimmchip/.style={draw, fill=black!85, minimum width=0.3cm, minimum height=0.24cm, inner sep=0pt},
      dimmpcb/.style={draw, very thick, fill=white},
      arr/.style={-{Latex[length=1.4mm]}, thick, line cap=round},
      lbl/.style={font=\sffamily\footnotesize},
      sublbl/.style={font=\sffamily\scriptsize}
    ]

    % ========================================================
    %  LEFT PANEL: Intel Arrow Lake iMC
    % ========================================================
    \node[cpu, minimum width=2.3cm] (intelcpu) at (0, 0) {};
    \node[anchor=north, font=\sffamily\small, align=center] at ($(intelcpu.north)+(0,-0.1)$) {Intel Arrow\\Lake CPU};
    \node[mc, anchor=south] (intelimc) at ($(intelcpu.south)+(0,0.1)$) {iMC};

    \node[slot] (intelB) at (3.5, 0) {};
    \node[slot] (intelA) at (4.2, 0) {};
    \node[lbl, below=2pt of intelB] {(B)};
    \node[lbl, below=2pt of intelA] {(A)};

    % CMC0: route up-and-right to upper region, then split to both slots
    \draw[arr] ($(intelimc.north)+(-0.1,0)$) -- ($(intelimc.north)+(-0.1,0.2)$) -- ($(intelimc.north)+(0.7,0.2)$) coordinate (c0fan);
    \draw[arr] (c0fan) -| ($(intelB.west)+(0.2,0.7)$);
    \draw[arr] (c0fan) -| ($(intelA.west)+(0.2,0.7)$);
    \node[lbl, anchor=south] at ($(c0fan)+(1.2,0.05)$) {CMC0};
    \node[sublbl] at ($(intelB.west)+(-0.3,0.8)$) {SC1};
    \node[sublbl] at ($(intelA.west)+(-0.1,0.8)$) {SC0};

    % CMC1: route down-and-right to lower region, then split
    \draw[arr] ($(intelimc.south)+(-0.1,0)$) -- ($(intelimc.south)+(-0.1,-0.4)$) -- ($(intelimc.south)+(0.7,-0.4)$) coordinate (c1fan);
    \draw[arr] (c1fan) -| ($(intelB.west)+(0.2,-0.7)$);
    \draw[arr] (c1fan) -| ($(intelA.west)+(0.2,-0.7)$);
    \node[lbl, anchor=north] at ($(c1fan)+(1.2,-0.05)$) {CMC1};
    \node[sublbl] at ($(intelB.west)+(-0.3,-0.6)$) {SC1};
    \node[sublbl] at ($(intelA.west)+(-0.1,-0.6)$) {SC0};

    % DIMM
    \node[dimmpcb, minimum height=2.0cm, minimum width=1.0cm] (inteldimm) at (5.4, 0) {};
    \foreach \y in {0.65, 0.35, 0.05} {\node[dimmchip] at (5.4, \y) {};}
    \foreach \y in {-0.05, -0.35, -0.65} {\node[dimmchip] at (5.4, \y) {};}
    \node[lbl, anchor=west] at ($(inteldimm.east)+(0.08,0.4)$) {32 bit};
    \node[lbl, anchor=west] at ($(inteldimm.east)+(0.08,-0.4)$) {32 bit};
    \draw[thin] ($(inteldimm.east)+(0,0.78)$) -- ++(0.07,0);
    \draw[thin] ($(inteldimm.east)+(0,0.02)$) -- ++(0.07,0);
    \draw[thin] ($(inteldimm.east)+(0,-0.02)$) -- ++(0.07,0);
    \draw[thin] ($(inteldimm.east)+(0,-0.78)$) -- ++(0.07,0);

    \node[lbl, below=0.65cm of intelA, xshift=-0.35cm] {Memory DIMMs};
    \node[font=\sffamily\bfseries\small] at (3.0, -2.3) {(a) Intel iMC (Arrow Lake)};

    % ========================================================
    %  RIGHT PANEL: AMD UMC
    % ========================================================
    \begin{scope}[xshift=8.4cm]
      \node[cpu, minimum width=2.3cm] (amdcpu) at (0, 0) {};
      \node[anchor=north, font=\sffamily\small, align=center] at ($(amdcpu.north)+(0,-0.1)$) {Granite Ridge\\AM5 CPU};
      \node[mc, anchor=south] (amdumc) at ($(amdcpu.south)+(0,0.1)$) {UMC};

      \node[slot] (amdB) at (3.5, 0) {};
      \node[slot] (amdA) at (4.2, 0) {};
      \node[lbl, below=2pt of amdB] {(B)};
      \node[lbl, below=2pt of amdA] {(A)};

      \draw[arr] ($(amdumc.north)+(-0.1,0)$) -- ($(amdumc.north)+(-0.1,0.2)$) -- ++(0.7,0) coordinate (chamid) -| ($(amdA.west)+(0.2,0.2)$);
      \node[lbl, anchor=south] at ($(chamid)+(1.2,0.05)$) {CHA};

      \draw[arr] ($(amdumc.south)+(-0.1,0)$) -- ($(amdumc.south)+(-0.1,-0.35)$) -- ++(0.7,0) coordinate (chbmid) -| ($(amdB.west)+(0.2,-0.45)$);
      \node[lbl, anchor=north] at ($(chbmid)+(1.2,-0.15)$) {CHB};

      \node[dimmpcb, minimum height=2.0cm, minimum width=1.0cm] (amddimm) at (5.4, 0) {};
      \foreach \y in {0.75, 0.45, 0.15, -0.15, -0.45, -0.75} {\node[dimmchip] at (5.4, \y) {};}
      \node[lbl, anchor=west] at ($(amddimm.east)+(0.08,0)$) {64 bit};
      \draw[thin] ($(amddimm.east)+(0,0.9)$) -- ++(0.07,0);
      \draw[thin] ($(amddimm.east)+(0,-0.9)$) -- ++(0.07,0);

      \node[lbl, below=0.65cm of amdA, xshift=-0.35cm] {Memory DIMMs};
      \node[font=\sffamily\bfseries\small] at (3.0, -2.3) {(b) AMD UMC (AM5)};
    \end{scope}

    % ========================================================
    %  BOTTOM ROW
    % ========================================================
    \begin{scope}[yshift=-3.4cm]
      \node[dimmpcb, minimum width=4.4cm, minimum height=0.8cm] (regdimm) at (3.5, 0) {};
      \foreach \x in {-1.55, -1.2, -0.85, -0.5} {\node[dimmchip] at ($(regdimm.center)+(\x,0)$) {};}
      \foreach \x in {0.5, 0.85, 1.2, 1.55} {\node[dimmchip] at ($(regdimm.center)+(\x,0)$) {};}
      \node[lbl, below=2pt of regdimm, font=\sffamily\bfseries\footnotesize] {Regular DDR5 module};

      \node[dimmpcb, minimum width=4.4cm, minimum height=0.8cm] (scdimm) at (10.0, 0) {};
      \foreach \x in {-1.55, -1.2, -0.85, -0.5} {\node[dimmchip] at ($(scdimm.center)+(\x,0)$) {};}
      \node[lbl, below=2pt of scdimm, font=\sffamily\bfseries\footnotesize] {Single 32-bit sub-channel DDR5 module};
    \end{scope}

  \end{tikzpicture}
  \caption{Intel iMC vs.\ AMD UMC sub-channel topology, and corresponding standard vs.\ single sub-channel module physical realisations. (a) On Intel Arrow Lake, the iMC contains two memory controllers (CMC0, CMC1); each CMC drives two physically separate DIMM slots, with the two 32-bit sub-channels (SC0, SC1) of a given CMC routed to different slots. A standard DDR5 DIMM in either slot exposes a 32-bit interface to each of the two CMCs. (b) On AMD AM5, a single UMC drives two channels (CHA, CHB), each as a unified 64-bit interface to one DIMM slot. Bottom: a standard module populates both 32-bit halves with DRAM dies; a single sub-channel module populates only one half.}
  \label{fig:imc_umc}
\end{figure*}

The two CPU vendors with consumer DDR5 platforms in 2026, Intel and AMD, take fundamentally different approaches to channel and sub-channel organisation. Figure~\ref{fig:imc_umc} illustrates the two topologies side by side as realised on current platforms (Intel Arrow Lake, AMD Granite Ridge / AM5). The architectural choice in each case has direct consequences for whether asymmetric (single sub-channel) module population is supported.

\subsection{Intel Integrated Memory Controller}

Intel's DDR5 iMC, introduced with Alder Lake (12th Gen Core, 2021) \cite{intel_ddr5_2021}, treats each 32-bit sub-channel as an independent scheduling domain with its own command queue, timing state machine, and PHY instance. This independence is preserved in the Arrow Lake topology shown in Figure~\ref{fig:imc_umc}(a) even though the physical sub-channel routing has changed across Intel platform generations (Section~\ref{sec:slot_topology}). At initialisation the iMC reads SPD \cite{jedec_spdpub} to detect per-sub-channel population, skips training and calibration for absent sub-channels, and routes traffic only through populated ones. Single sub-channel operation is therefore a degenerate case of the iMC's standard operating model rather than an exceptional code path.

\subsection{AMD Unified Memory Controller}

AMD's Zen 4 UMC \cite{amd_zen4_2022}, introduced with AM5 (2022), implements each DDR5 channel as a unified 64-bit logical entity (Figure~\ref{fig:imc_umc}(b)). Training algorithms and initialisation sequences --- read levelling, write levelling, DQ skew calibration --- calibrate all 64 data lines of a channel as a coordinated group. While JEDEC DDR5 \cite{jedec_ddr5_2020} defines on-die termination (ODT) for populated DQ lanes, the DQ lines corresponding to an unpopulated 32-bit half of a DIMM present an unterminated stub at the connector. The resulting impedance mismatch and signal-integrity violations during the UMC's coordinated 64-bit training sequence cause training assertions to fail and abort POST. This is an architectural constraint inherent to the UMC design philosophy, not a firmware issue amenable to simple patching.

\subsection{Architectural Dichotomy}

The fundamental design difference is that the Intel iMC schedules independently at the 32-bit sub-channel granularity, while the AMD UMC schedules at the 64-bit channel granularity with internal bank-group parallelism. Both are valid design choices reflecting different historical trade-offs between scheduling flexibility and training simplicity. Table~\ref{tab:imc_umc} summarises the key controller differences.

\begin{table}[!htbp]
  \centering
  \caption{Intel iMC vs. AMD UMC: Sub-Channel Handling}
  \label{tab:imc_umc}
  \small
  \setlength{\tabcolsep}{4pt}
  \begin{tabular}{@{}lll@{}}
    \toprule
    \textbf{Property} & \textbf{Intel iMC} & \textbf{AMD UMC} \\
    \midrule
    Scheduling granularity & 32-bit sub-channel & 64-bit channel \\
    Single SC support      & Yes               & No \\
    Training scope         & Per sub-channel   & All 64-bit \\
    Asymm. population      & Supported         & Boot failure \\
    PHY power gating       & Per sub-channel   & Full channel \\
    SPD SC detection       & Used at init      & N/A \\
    \bottomrule
  \end{tabular}
\end{table}

The AMD architectural approach bounds the addressable deployment range for single sub-channel DDR5 modules to Intel platforms. AMD has not announced support for the configuration. A firmware path is conceivable in principle --- for example, modifying AGESA to skip training assertions in the presence of an SPD flag, or shipping motherboards with dummy load circuits during training --- but such modifications would diverge from the unified UMC training model on which AM5 platform validation is currently based, and would require coordinated AGESA changes outside the scope of any single motherboard vendor. We therefore characterise AMD AM5 as outside the addressable range for this module class for the lifetime of the platform.

% ── VII. Module Design and System Implications ────────────────────────────
\section{Module Design and System Implications}

\subsection{Die Count and Capacity}

A single sub-channel single-rank consumer UDIMM in $\times$8 organisation requires four data dies:

\begin{equation}
C_{\min} = N_{\text{dies}} \times D_{\text{die}} = 4 \times 16\,\text{Gb} = 64\,\text{Gb} = 8\,\text{GB}.
\label{eq:cmin}
\end{equation}

Table~\ref{tab:dimm_compare} presents the comparative module specifications.

\begin{table}[!htbp]
  \centering
  \caption{Standard vs. Single Sub-Channel DDR5 DIMM}
  \label{tab:dimm_compare}
  \small
  \setlength{\tabcolsep}{4pt}
  \begin{tabular}{@{}lll@{}}
    \toprule
    \textbf{Parameter} & \textbf{Standard} & \textbf{Single SC} \\
    \midrule
    Active sub-channels       & 2 (SC0+SC1)    & 1 (SC0 only) \\
    Data width                & 64-bit         & 32-bit \\
    Data dies ($\times$8, 16\,Gb) & 8          & 4 \\
    Min.\ capacity            & 16\,GB         & 8\,GB \\
    BOM reduction             & ---            & 35--45\% \\
    SC1 data pins             & Active         & NC/tied \\
    Power (typ.)              & 100\%          & vendor-specific  \\
    \bottomrule
  \end{tabular}
\end{table}

\subsection{PCB and Electrical Considerations}

Halving the data bus width has cascading PCB implications:

\begin{itemize}
  \item \textbf{Trace count}: 32 DQ traces instead of 64 (plus corresponding DQS pairs); significant routing simplification.
  \item \textbf{PCB layers}:  Halving the data trace count opens the possibility of layer-count reduction in module PCB design, particularly for SO-DIMM form factors. The exact reduction is vendor- and speed-grade-specific.
  \item \textbf{Signal integrity margin}: Reducing the populated-DQ trace count lowers aggressor density on the routed half of the module. Whether this translates into measurable margin improvement at a given speed grade depends on channel geometry and is module-design specific.
  \item \textbf{Termination}: The unpopulated sub-channel's DQ pins present an unterminated stub at the DIMM connector. Module designers must define a termination strategy (no-connect, pull-down, or terminator-resistor) and validate eye diagrams at target data rates on the populated sub-channel, particularly at the upper end of the consumer DDR5 envelope (5600--6400\,MT/s).
  \item \textbf{Decoupling}: SC1 traces carrying no signal cannot act as aggressors against SC0; routine within-sub-channel cross-coupling rules apply unchanged.
\end{itemize}

\subsection{Firmware Requirements}

No motherboard PCB redesign is required. BIOS/UEFI firmware updates must:
\begin{enumerate}
  \item Parse SPD bytes to identify single sub-channel modules \cite{jedec_spdpub};
  \item Skip Sub-Channel B training phases;
  \item Adjust the OS-visible memory address map to reflect reduced channel width;
  \item Adapt XMP profile parsing.
\end{enumerate}

Power delivery requirements are reduced because half the DRAM die population draws no operating current. The exact reduction depends on workload, refresh rate, and PMIC overhead, and is module-design specific.

\subsection{CPU Memory Controller Implications}

For CPU designers, single sub-channel support primarily impacts the memory controller PHY. A full dual sub-channel DDR5 channel requires two complete PHY instances per memory channel. Supporting single sub-channel mode requires the ability to power-gate or bypass one PHY, which modern Intel designs implement via per-channel power gating. A single DDR5 PHY at advanced nodes (Intel 4 / 7\,nm-class) occupies approximately 2--4\,mm$^2$. Designs shipping single sub-channel only (e.g., some embedded SKUs) could realise meaningful area and power savings; for mobile platforms where thermal envelope is critical, the reduced quiescent power of an idled sub-channel is non-trivial.

% ── VIII. Cost and Deployment Analysis ────────────────────────────────────
\section{Cost and Deployment Analysis}
\label{sec:market}

\subsection{DRAM Supply Context}

The configuration analysed in this paper has been technically feasible since the introduction of DDR5 in 2020 \cite{jedec_ddr5_2020} and has been used in extreme operating conditions since 2021 (Section~\ref{sec:oc_priority}). Its productisation as a cost-reduced volume module class in 2025--2026 is associated with two contemporaneous shifts in DRAM supply. First, AI-related demand for HBM and high-density server DDR5 has led the major three memory manufacturers (Samsung, SK hynix, Micron) to reallocate substantial leading-edge wafer capacity toward those product categories, with HBM consuming approximately three times the wafer area per bit of equivalent commodity DDR5 \cite{tomshw_hbm_2025} and AI-related DRAM consumption projected at approximately 20\% of global wafer output in 2026 \cite{trendforce_ai_dram_2025}. Industry forecasts indicate that supply tightness is likely to persist through 2027, with some analyses extending into 2028 \cite{taipeitimes_nanya_2026, idc_memory_2026}. Second, DDR4 production has contracted concurrently: by Q4 2026, global DDR4 capacity is forecast to fall to 25--33\% of Q1 2025 levels \cite{trendforce_ddr4_2025}, with residual supply concentrated in a small set of Taiwan-based vendors (Nanya, Winbond, Powerchip) \cite{digitimes_ddr4_2025, trendforce_ddr4_2025}. The 8\,GB tier customer in this period faces a different cost calculus than in previous DRAM cycles: DDR4 no longer functions as a stable cost baseline against which DDR5 alternatives can be evaluated.

\subsection{Bill-of-Materials Decomposition}

Absolute price estimates under current supply conditions are unstable, since spot prices have fluctuated on weekly timescales \cite{tomshw_ram_index_2026}. We therefore characterise the BOM \emph{ratio} between standard and single sub-channel modules at a given die generation, which is invariant to absolute price levels. Assuming pre-supply-constraint pricing of \$3.50--\$4.50 per DDR5-5600 16\,Gbit $\times$8 die \cite{micron_ddr5_2022}, a standard 16\,GB single-rank consumer UDIMM (8 dies, dual-SC PCB)  corresponds to approximately \$30--\$40 BOM, and a single sub-channel 8\,GB DDR5 module (4 dies, simplified PCB) corresponds to approximately \$15--\$20 BOM. The ratio of approximately 50\% module BOM at half the capacity, or equivalently a 35--45\% saving when normalised against fixed PCB and assembly costs, is the structurally invariant figure.

Table~\ref{tab:bom_breakdown} decomposes the BOM saving by component. DRAM die count is the dominant contributor: dies constitute approximately two-thirds of standard module BOM, and halving the die count accounts for most of the total saving. PCB layer-count reduction provides a smaller secondary contribution. The on-DIMM PMIC is essentially unchanged between configurations, since a single regulator services the module regardless of sub-channel population.

\begin{table}[!htbp]
  \centering
  \caption{BOM Reduction Breakdown by Component (Single SC vs. Standard DDR5)}
  \label{tab:bom_breakdown}
  \small
  \setlength{\tabcolsep}{4pt}
  \begin{tabular}{@{}lcc@{}}
    \toprule
    \textbf{Component} & \textbf{\% of std.\ BOM} & \textbf{Single-SC saving} \\
    \midrule
    DRAM dies & 60--70\% & $\approx$50\% \\
    PCB (substrate + layers) & 15--20\% & $\approx$30--40\% \\
    PMIC                  & 5--10\% & negligible \\
    Passives + assembly   & 5--10\% & $\approx$20\% \\
    \midrule
    \textbf{Total module BOM} & \textbf{100\%} & \textbf{$\approx$35--45\%} \\
    \bottomrule
  \end{tabular}
\end{table}

\subsection{Die Density Dependence}

The capacity argument depends on the prevailing die density. With 16\,Gbit dies, a standard DIMM requires 8 data dies to reach 16\,GB and a single sub-channel module requires 4 data dies to reach 8\,GB. With 32\,Gbit dies (anticipated to reach volume production in 2026--2027 in pre-supply-constraint roadmaps), the corresponding die counts become 4 and 2. The relative BOM advantage of the single sub-channel configuration scales with the absolute die-count differential:
\begin{equation}
\Delta_{\text{BOM}}(t) \propto \frac{N_{\text{std}}(t) - N_{\text{SC}}(t)}{N_{\text{std}}(t)}
\end{equation}
where $N(t)$ denotes die count per capacity tier at time $t$. The 50\% die-count ratio is preserved through the 32\,Gbit transition, Any PCB-layer savings would diminish at lower die counts, since module routing complexity is reduced even at the standard configuration. The cost advantage is therefore most meaningful during the 16\,Gbit-die generation; the 32\,Gbit transition narrows it.

\subsection{Deployment Suitability}

The bandwidth, platform, and cost analyses jointly delineate the deployment range. Single sub-channel DDR5 is appropriate for: cost-sensitive Intel desktop and SO-DIMM platforms without a discrete GPU; Chromebook-class and education devices; thin-client, kiosk, and embedded platforms on Intel-derived SoCs; and OEM 8\,GB-tier systems where the alternative DDR4 sourcing path is unreliable. It is not appropriate for: gaming systems with high-framerate workloads; iGPU-dependent laptops with 1080p or higher sustained graphics targets; workstation and CPU-side AI inference platforms; or any AMD AM5 platform. The boundary between these classes is determined by the joint constraint of platform memory controller architecture (Section~\ref{sec:platform}) and workload bandwidth sensitivity (Section~\ref{sec:perf}), both of which are observable at design time.

% ── IX. JEDEC Standardisation Proposal ────────────────────────────────────
\section{The JEDEC Encoding and the Standardisation Gap Around It}
\label{sec:jedec}

\subsection{JEDEC Already Defines the SPD Field}
\label{sec:jedec_spd_field}

A common assumption in industry discussion of cost-reduced DDR5 modules is that single sub-channel population is an unstandardised configuration requiring vendor-specific SPD extensions. The JEDEC SPD specification for DDR5 modules, JESD400-5D.01 \cite{jedec_spdpub}, in fact already defines the sub-channel count as a per-DIMM SPD field. Section 11.11, ``(Common): Memory Channel Bus Width,'' defines Byte 235 (offset 0xEB), Table 104, with the following encoding:

\begin{itemize}
  \item \textbf{Bits 7--5: Number of Sub-Channels per DIMM.} \texttt{000} = 1 sub-channel; \texttt{001} = 2 sub-channels (the existing default for standard UDIMMs/SO-DIMMs); \texttt{010} = 4 sub-channels; \texttt{011} = 8 sub-channels; all other values reserved.
  \item Bits 4--3: Bus extensions per sub-channel (ECC width).
  \item Bits 2--0: Primary bus width per sub-channel (8 / 16 / 32 / 64 bits).
\end{itemize}

A 32-bit single sub-channel module is therefore expressible \textit{today} within the existing JEDEC SPD encoding: bits 7--5 = \texttt{000}, bits 2--0 = \texttt{010}. This is not an inferred reading of the standard but the literal text of Table 104 in JESD400-5D.01. Every world record discussed in Section~\ref{sec:oc_priority} was set on a configuration whose SPD content is fully JEDEC-conformant under this encoding, and a host iMC encountering this SPD content has, in principle, all the information it needs to initialise the module correctly. The JEDEC standard explicitly anticipated this configuration; it did not require a standards revision to introduce it.

\subsection{The Real Standardisation Gap}

The gap is therefore not at the SPD-field level but at the surrounding ecosystem layers. Five specific deficiencies mean that a SPD-conformant single sub-channel module cannot today reliably be brought to market as a standard interoperable product:

\textbf{(1) Termination specification.} The base DDR5 standard \cite{jedec_ddr5_2020} defines on-die termination for populated DQ lanes but does not specify required termination behaviour for the unpopulated SC1 DQ, DQS, and CA pins on a single sub-channel module. Practice varies between leaving pins as no-connect, tying to ground, and terminating to $V_{DDQ}$/2 via on-module resistors; signal-integrity outcomes differ across these choices, but no JEDEC document mandates a particular approach or specifies eye-diagram acceptance margins for the host-side receiver under canonical channel topologies.

\textbf{(2) Platform compatibility signalling.} A SPD-conformant single sub-channel module inserted into an AMD AM5 system today will produce a silent training failure during POST (Section~\ref{sec:platform}). No SPD bit indicates platform compatibility class (Intel iMC compatible / AMD UMC compatible / both), so the module cannot self-identify as incompatible with the host before training. An asymmetry-aware compatibility flag would allow host platforms to refuse incompatible modules cleanly with a diagnostic message rather than fail mid-training.

\textbf{(3) Module nomenclature.} JEDEC has no standardised part-number suffix or labelling convention for single sub-channel modules. In the absence of a JEDEC convention, module vendors and motherboard vendors have begun to introduce competing trade names for the same underlying configuration; this fragmentation is not in the long-term interest of the ecosystem and is precisely what JEDEC standardisation is meant to prevent. A standardised JEDEC suffix (analogous to existing \texttt{-R} for registered or \texttt{-E} for ECC) would allow module identification independent of vendor branding.

\textbf{(4) Performance-profile extension.} Existing XMP and EXPO profile schemas assume both sub-channels are populated. Single sub-channel modules wishing to publish optimised timings for the SC0-only operating mode require schema extensions that allow per-sub-channel timing sets without breaking dual-SC profile parsers on legacy platforms.

\textbf{(5) Validation test procedures.} JEDEC has not published a compliance test suite for single sub-channel modules, covering: SPD parsing correctness; iMC training behaviour with and without optional SC1 dummy-load circuits; eye-diagram acceptance margins at platform target speeds; and interoperability across the DIMM-slot population matrix (1 SC1 module + 1 standard module, 2 SC1 modules, 1 SC1 module alone). A reference-platform validation kit would allow module vendors to qualify against a defined Intel platform target without per-vendor BIOS variation.

\subsection{Proposal Summary}

The technical content needed to close this gap is small relative to the size of the existing JEDEC DDR5 standards corpus: an SPD termination-scheme byte, a platform-compatibility bit, a part-number suffix convention, an XMP/EXPO schema extension, and a validation-test annex. The work is well within the scope of a single JEDEC working-group cycle. We propose that this work be undertaken with reference to the existing single sub-channel module class designated by SPD Byte 235 bits 7--5 = \texttt{000}, with the JEDEC adoption preserving alignment to the encoding already in the standard rather than introducing a parallel encoding scheme. For convenience in this paper we refer to such a module as \textbf{DDR5-SC1}.

\subsection{Adoption Path}

Closing this gap requires a small, bounded amount of additional standardisation activity relative to the existing JEDEC DDR5 corpus, well within the scope of a single working-group cycle. Vendor-specific implementations introduced ahead of formal standardisation may diverge in choices of termination, module nomenclature, and platform-compatibility signalling, in which case the role of the JEDEC standard is to consolidate the divergent choices and reduce the qualification burden on subsequent entrants. We do not attempt to predict adoption timing.

% ── X. Discussion ─────────────────────────────────────────────────────────
\section{Discussion}

\subsection{Limitations}

This work has the following limitations.

\textbf{Performance estimates are model-derived.} The performance deficit ranges in Table~\ref{tab:workloads} are obtained from roofline-model projections and have not been validated against silicon measurements. Real-world deviations may arise from memory controller prefetch behaviour, LLC capacity effects, and workload-specific access patterns. We treat the model results as upper bounds on bandwidth-driven deficit and lower bounds on latency-driven deficit, but precise per-workload numbers will require measurement on production hardware.

\textbf{Platform analysis is documentation-based.} The AMD incompatibility assessment is based on publicly disclosed architectural documentation \cite{amd_zen4_2022} and on the absence of vendor announcements regarding asymmetric population support. Future UMC revisions could in principle add such support; the analysis here describes the current state of the platform.

\textbf{Cost analysis is sensitive to supply conditions.} BOM ratios are stable across pricing regimes by construction, but absolute cost estimates assume pre-supply-constraint die pricing. The configuration's competitive position relative to standard DDR5 and to DDR4 is sensitive to the trajectory of DRAM supply normalisation.

\subsection{Future Work}

Three directions warrant further investigation:
\begin{enumerate}
  \item Empirical validation of the performance estimates on production single sub-channel SO-DIMM hardware on Intel Lunar Lake or Arrow Lake platforms, including signal-integrity measurements (eye diagrams) at 5600--7200\,MT/s on production-representative motherboard topologies.
  \item Engagement with the JEDEC working group to close the ecosystem standardisation gap around the existing SPD Byte 235 sub-channel-count encoding, along the lines outlined in Section~\ref{sec:jedec}.
  \item Analysis of the LPCAMM2 form factor, where per-sub-channel power gating may provide power-efficiency benefits beyond the BOM saving discussed here.
\end{enumerate}

% ── XI. Conclusion ────────────────────────────────────────────────────────
\section{Conclusion}

We have presented an architectural and performance analysis of the 32-bit single sub-channel DDR5 module configuration. The mathematical identity $32\,\text{bit} \times BL16 = 64\,\text{bytes} = L_{x86}$, embedded in the JEDEC DDR5 design, shows that a single sub-channel transfers one full x86 cache line per burst transaction, and we have shown that the configuration is well-characterised electrically through its use in HWBOT-verified frequency records on three successive Intel platform generations. The configuration is therefore not a new architecture in 2026 but a use-case repositioning of an existing technique toward cost-reduced volume modules under contemporaneous DRAM supply pressure.

The trade-offs that bound the configuration's deployment range are quantifiable: a 50\% peak bandwidth reduction relative to standard DDR5, a bandwidth inversion against DDR4-3200 at the DDR5-4800 tier, insufficient bandwidth headroom for sustained iGPU workloads at 1080p and above, and incompatibility with AMD AM5 platforms as a direct consequence of the unified UMC training model. The BOM reduction of 35--45\% relative to standard DDR5 is meaningful primarily during the 16\,Gbit-die generation, with the 32\,Gbit transition narrowing the absolute differential. Finally, we have shown that JEDEC's SPD specification (JESD400-5D.01) already encodes the single sub-channel module class in Byte 235, and identified the surrounding ecosystem standardisation work --- termination, platform-compatibility signalling, module nomenclature, validation --- as the substantive remaining task for formal interoperability across the ecosystem.

% ── References ────────────────────────────────────────────────────────────
\printbibliography

@misc{jedec_ddr5_2020,
  author       = {{JEDEC Solid State Technology Association}},
  title        = {{DDR5 SDRAM Standard JESD79-5B}},
  howpublished = {JEDEC Publication},
  year         = {2020},
  url          = {https://www.jedec.org/standards-documents/docs/jesd79-5b}
}

@misc{jedec_ddr4_2020,
  author       = {{JEDEC Solid State Technology Association}},
  title        = {{DDR4 SDRAM Standard JESD79-4C}},
  howpublished = {JEDEC Publication},
  year         = {2020},
  url          = {https://www.jedec.org/standards-documents/docs/jesd79-4c}
}

@misc{jedec_spdpub,
  author       = {{JEDEC Solid State Technology Association}},
  title        = {{Serial Presence Detect (SPD), General Standard for Memory Module Types --- DDR5 SDRAM (JESD400-5D.01)}},
  howpublished = {JEDEC Publication No.\ 400-5D.01},
  year         = {2024},
  note         = {Successor to JESD21C Annex L for the DDR5 generation. Section 11.11 ``(Common): Memory Channel Bus Width'' defines Byte 235 (offset 0xEB), Table 104, with bits 7--5 encoding the number of sub-channels per DIMM (000 = 1 sub-channel, 001 = 2 sub-channels, 010 = 4 sub-channels, 011 = 8 sub-channels)}
}

@techreport{intel_ddr5_2021,
  author       = {{Intel Corporation}},
  title        = {{12th Generation Intel Core Processors Datasheet, Volume 1 of 2}},
  institution  = {Intel Corporation},
  number       = {Doc. No. 655258},
  type         = {Datasheet},
  year         = {2021},
  url          = {https://edc.intel.com/content/www/us/en/design/ipla/software-development-platforms/client/platforms/alder-lake-desktop/12th-generation-intel-core-processors-datasheet-volume-1-of-2/}
}

@techreport{amd_zen4_2022,
  author       = {{Advanced Micro Devices}},
  title        = {{Software Optimization Guide for the AMD Zen4 Microarchitecture}},
  institution  = {Advanced Micro Devices},
  number       = {Publication \#57647, Rev. 1.00},
  type         = {Technical Reference},
  year         = {2023},
  month        = jan
}

@book{patterson_hennessy_2020,
  author    = {Patterson, David A. and Hennessy, John L.},
  title     = {Computer Organization and Design: ARM Edition},
  edition   = {2nd},
  publisher = {Morgan Kaufmann},
  year      = {2020},
  chapter   = {5}
}

@inproceedings{malladi2012dram,
  author    = {Malladi, Krishna T. and Lee, Benjamin C. and Nothaft, Frank A. and Kozyrakis, Christos and Periyathambi, Karthik and Horowitz, Mark},
  title     = {Towards energy-proportional datacenter memory with mobile {DRAM}},
  booktitle = {Proceedings of the 39th Annual International Symposium on Computer Architecture (ISCA)},
  address   = {Portland, OR, USA},
  pages     = {37--48},
  year      = {2012},
  doi       = {10.1109/ISCA.2012.6237004}
}

@techreport{micron_ddr5_2022,
  author       = {Rooney, Randall and Koyle, Neal},
  title        = {{Micron DDR5 SDRAM: New Features}},
  institution  = {Micron Technology, Inc.},
  number       = {CCM004-676576390-11390, Rev.\ A},
  type         = {White Paper},
  year         = {2019},
  month        = nov
}

@misc{trendforce_ai_dram_2025,
  author       = {{TrendForce}},
  title        = {{AI Reportedly to Consume 20\% of Global DRAM Wafer Capacity in 2026, HBM and GDDR7 Lead Demand}},
  howpublished = {TrendForce News},
  year         = {2025},
  month        = dec,
  url          = {https://www.trendforce.com/news/2025/12/26/news-ai-reportedly-to-consume-20-of-global-dram-wafer-capacity-in-2026-hbm-gddr7-lead-demand/}
}

@misc{idc_memory_2026,
  author       = {Jeronimo, Francisco},
  title        = {{Global Memory Shortage Crisis: Market Analysis and the Potential Impact on the Smartphone and PC Markets in 2026}},
  howpublished = {IDC Research Blog},
  year         = {2026},
  month        = feb,
  url          = {https://www.idc.com/resource-center/blog/global-memory-shortage-crisis-market-analysis-and-the-potential-impact-on-the-smartphone-and-pc-markets-in-2026/}
}

@misc{tomshw_hbm_2025,
  author       = {{Tom's Hardware}},
  title        = {{HBM is Coming for Your PC's RAM: HBM Consumes Around Three Times the Wafer Capacity of DDR5}},
  howpublished = {Tom's Hardware},
  year         = {2025},
  month        = dec,
  url          = {https://www.tomshardware.com/pc-components/ram/hbm-is-eating-your-ram}
}

@misc{tomshw_ram_index_2026,
  author       = {{Tom's Hardware}},
  title        = {{RAM Price Tracking 2026: Daily Lowest Price on DDR5 and DDR4 Memory During the AI-Driven Pricing Crisis}},
  howpublished = {Tom's Hardware Price Index},
  year         = {2026},
  url          = {https://www.tomshardware.com/pc-components/ram/ram-price-index-2026-lowest-price-on-ddr5-and-ddr4-memory-of-all-capacities}
}

@misc{trendforce_ddr4_2025,
  author       = {{TrendForce}},
  title        = {{Taiwan DRAM Makers Reportedly Eye 20--50\% Q4 Contract Price Hikes amid DDR4 Supply Squeeze}},
  howpublished = {TrendForce News},
  year         = {2025},
  month        = sep,
  url          = {https://www.trendforce.com/news/2025/09/11/news-taiwan-dram-makers-reportedly-eye-20-50-q4-contract-price-hikes-amid-ddr4-supply-squeeze/}
}

@misc{taipeitimes_nanya_2026,
  author       = {{Taipei Times}},
  title        = {{DRAM Shortage to Last Through 2028: Nanya Technology}},
  howpublished = {Taipei Times},
  year         = {2026},
  month        = mar,
  url          = {https://www.taipeitimes.com/News/biz/archives/2026/03/05/2003853263}
}

@misc{digitimes_ddr4_2025,
  author       = {{DigiTimes}},
  title        = {{Who Controls DDR4 Now? Nanya Steps in as Global Giants Shift Focus}},
  howpublished = {DigiTimes Asia},
  year         = {2025},
  month        = jul,
  url          = {https://www.digitimes.com/news/a20250728PD222/ddr4-market-capacity-winbond-ddr5.html}
}

@misc{gigabyte_ddr5_10022_2022,
  author       = {{GIGABYTE Technology}},
  title        = {{DDR5-10022 World Record Set! Overclocking with Z690 AORUS TACHYON}},
  howpublished = {GIGABYTE Press Release},
  year         = {2022},
  month        = may,
  note         = {Memory frequency 5011\,MHz on Intel Core i9-12900K (Alder Lake), single 32-bit sub-channel configuration; HWBOT submission 5379, set by overclocker HiCookie},
  url          = {https://www.gigabyte.com/bz/Press/News/1988}
}

@misc{hwbot_5379_2022,
  author       = {{HiCookie}},
  title        = {{HWBOT Memory Frequency World Record: DDR5 SDRAM 5011\,MHz (DDR5-10022)}},
  howpublished = {HWBOT Submission \#5379},
  year         = {2022},
  month        = may,
  url          = {https://hwbot.org/newsflash/5379_hicookie_sets_new_ddr5_memory_frequency_world_record_at_5011mhz/}
}

@misc{gigabyte_ddr5_11618_2023,
  author       = {{GIGABYTE Technology}},
  title        = {{Z790 AORUS TACHYON X Shatters Multiple World Records: DDR5-11618}},
  howpublished = {GIGABYTE Press Release},
  year         = {2023},
  month        = oct,
  note         = {Memory frequency 5809.2\,MHz on Intel Core i9-14900K (Raptor Lake Refresh), single 32-bit sub-channel configuration; HWBOT submission 5376447, set by HiCookie at IEM 2023 Sydney},
  url          = {https://www.gigabyte.com/bz/Press/News/2120}
}

@misc{hwbot_5376447_2023,
  author       = {{HiCookie}},
  title        = {{HWBOT Memory Frequency World Record: DDR5 SDRAM 5809.2\,MHz (DDR5-11618)}},
  howpublished = {HWBOT Submission \#5376447},
  year         = {2023},
  month        = oct,
  url          = {https://hwbot.org/submission/5376447_hicookie_memory_frequency_ddr5_sdram_5809.2_mhz/}
}

@misc{gigabyte_ddr5_13530_2025,
  author       = {{GIGABYTE Technology}},
  title        = {{Z890 AORUS TACHYON ICE Dominates Global DDR5 Performance, Shattering World Record at DDR5-13530}},
  howpublished = {GIGABYTE Press Release},
  year         = {2025},
  month        = nov,
  note         = {Memory frequency 6765\,MHz on Intel Core Ultra 200S (Arrow Lake), single 32-bit sub-channel configuration; HWBOT submission 5929126, set by Sergmann and HiCookie. Intel Chief Overclocking Architect Dan Ragland is quoted in the press release endorsing the result},
  url          = {https://www.gigabyte.com/bz/Press/News/2333}
}

@misc{hwbot_5929126_2025,
  author       = {{Sergmann and HiCookie}},
  title        = {{HWBOT Memory Frequency World Record: DDR5 SDRAM (DDR5-13530)}},
  howpublished = {HWBOT Submission \#5929126},
  year         = {2025},
  month        = nov,
  url          = {https://hwbot.org/benchmarks/memory_frequency/submissions/5929126}
}

@misc{intel_ddr5_13530_video_2025,
  author       = {{Intel Corporation}},
  title        = {{DDR5-13530 World Record on Intel Core Ultra 200S}},
  howpublished = {Intel-produced documentary video, featuring Dan Ragland (Intel Chief Overclocking Architect)},
  year         = {2025},
  url          = {https://www.youtube.com/watch?v=zbzR24TThrY}
}

\end{document}